\documentclass[12pt,preprint,runnignheads,natbib]{emulateapj}

\usepackage{graphicx}

\shorttitle{Impact of Sunspots on Coronal O VI}

\shortauthors{Morgan \& Habbal}

\begin{document}

\title{The Impact of Sunspots on the Interpretation of Coronal Observations of the O VI Doublet}

\author{H. Morgan\altaffilmark{1} and S. Rifai Habbal}

\affil{Institute for Astronomy, University of Hawaii, 2680 Woodlawn Drive, Honolulu, HI 96822, USA}

\altaffiltext{1}{hmorgan@ifa.hawaii.edu}

\begin{abstract}
Due to their high intensity of emission in the O VI 1031.9 and 1037.6\AA\ lines, even small sunspots on the solar disk can strongly influence the intensity of the radiative scattering component of O VI lines in the corona. Observations of O VI disk spectra show a 1032/1038 line intensity ratio of $>$2.6 in a sunspot compared to quiet disk values of $\sim$2. The enhancement of the 1032 line in comparison to the 1038 is likely due to interaction between molecular hydrogen emission from the sunspot and the chromospheric O$^{5+}$. Modeling shows that a contribution from sunspots increases the coronal O VI 1032/1038 intensity ratio to values considerably higher than those achieved with a quiet disk or coronal hole spectrum. Therefore a re-examination of flow velocities derived from UVCS/SOHO streamer observations must be made. This modeling demonstrates that the inclusion of sunspots, when present, may lead to non-zero outflow velocities at lower heights in streamer cores in contrast to some existing model results.
\end{abstract}

\section{INTRODUCTION}
\label{intro}

The radiative component of a coronal ultraviolet (UV) emission line is formed by photo-excitation of coronal ions by light from the Sun's disk and chromosphere. UV light from the disk and chromosphere consists of many high intensity emission lines and the coronal ions resonate with these so that the brightness of the radiative component of UV coronal lines is dependent on the disk and chromospheric emission \citep{gab1971a,gab1971b,bec1974}. The intensity of the radiative component of coronal UV lines is also sensitive to the velocity of the emitting ions relative to the Sun. UV coronal observations can therefore be used to place constraints on ion outflow velocities and temperatures \citep{hyd1970,wit1982}. 

Using Doppler dimming and pumping diagnostics, the intensity ratio of the O VI doublet lines at 1031.96 and 1037.60\AA\ (hereafter referred to as the 1032 and 1038 lines) can be used as evidence of bulk outflow velocity of O$^{5+}$ ions from the Sun \citep{koh1982, wit1982, noc1987,li1998}. The UltraViolet Coronagraph Spectrometer (UVCS) instrument aboard the Solar and Heliospheric Observatory (SOHO) is optimized for observations of the O VI 1032 and 1038 doublet \citep{koh1995}. Sophisticated empirical modeling of specific UVCS O VI observations have been used to constrain O$^{5+}$ outflow velocity and temperature parallel to the solar radial direction \citep[for example]{koh1997,str2002,fra2003}. 

Empirical models must employ a choice of solar disk incident radiation in order to calculate the radiative component of the coronal lines.  The general approach is to use a disk spectrum with a 1032/1038 line intensity ratio of around 2. This value is expected from a collisional plasma and agrees with measurements of the quiet Sun as described for example by \citet{noc1987}. This letter shows that if sunspots are present on the disk, their contribution to the disk radiation need to be included in empirical models. This is achieved through simple arguments in section \ref{diskspectra} based on observations of O VI emission from a sunspot. Section \ref{model}  examines the effect of a sunspot contribution on the coronal O VI intensity ratio as a function of outflow velocity using two simple coronal models - an isothermal spherically symmetric coronal hole and an axisymmetric streamer. Conclusions are given in section \ref{conclusions} along with an outlook for necessary future work.

\section{O VI DISK SPECTRA FROM SUNSPOTS}
\label{diskspectra}
\begin{figure}
\centering
\includegraphics[width=0.48\textwidth]{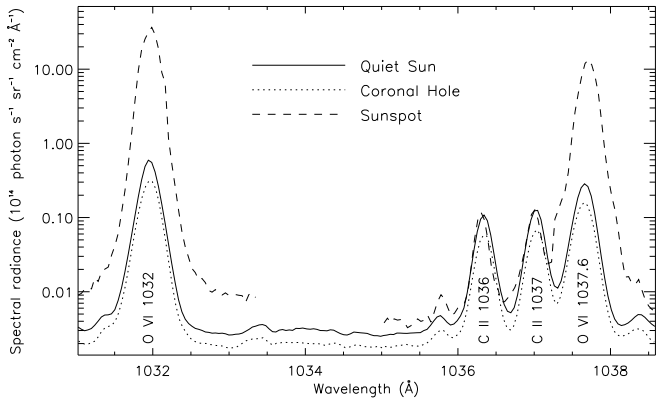}
		\caption{Disk spectra in the proximity of the O VI 1032 and 1038 doublet measured from quiet Sun, coronal hole and sunspot regions by SUMER, see \citet{cur2001}. Instrumental broadening of the lines is not compensated for here.}
\label{f1}
\end{figure}

A detailed study of UV disk spectra observed by the Solar Ultraviolet Measurement of Emitted Radiation (SUMER) instrument aboard SOHO is given by \citet{cur2001}. Of particular importance to the semi-empirical modeling of coronal UVCS observations is the disk spectrum in the proximity of the O VI doublet. Figure \ref{f1} shows the 1032 and 1038 disk spectra measured from quiet Sun, coronal hole and sunspot regions from data given by the SUMER solar atlas of \citet{cur2001}. Contained in the wavelength range are the two C II lines at 1037.0 and 1036.3\AA. These are the lines responsible for the Doppler pumping of the coronal O VI 1038 line \citep{koh1982,noc1987,li1998}. Table \ref{tb1} contains the peak intensities and linewidths obtained by fitting a background and a Gaussian to each line for all three spectra shown in figure \ref{f1}. For this table, instrumental broadening is removed from the linewidths using standard SUMER software contained in the Solar Software package. This software employs the SUMER instrumental widths determined by \citet{cha1998}. \citet{cur2001} gives a 1 $\sigma$ uncertainty in the radiometric calibration of SUMER as $\pm20\%$ for the quiet Sun and coronal hole measurements and $\pm30\%$ for the sunspot measurements. The increase in uncertainty is due to the sunspot observation being made during 1999 March 18, after the loss and recovery of SOHO's attitude control. These are very conservative estimates of the calibration uncertainties. 

\begin{table}[t]
\caption{Reduction of the spectra shown in figure \ref{f1}. $I$ is the peak intensity in $10^{13}$ photon s$^{-1}$ cm$^{-2}$ sr$^{-1}$ \AA$^{-1}$ and the $1/e$ halfwidths and $\lambda_{ref}$ are in \AA. Columns labeled QS, CH and SS refer to the quiet Sun, coronal hole and sunspot lines respectively. Conservative estimates of the uncertainty in intensities are 20\% for the quiet Sun and coronal hole, and 30\% for the sunspot.}
\begin{center}
\begin{tabular}{c|c|cc|cc|cc}
\multicolumn{2}{c}{} & \multicolumn{2}{c}{QS} & \multicolumn{2}{c}{CH}& \multicolumn{2}{c}{SS} \\
\hline
Line & $\lambda_{ref}$& $I$  & $\Delta\lambda$	& $I$  & $\Delta\lambda$ & $I$ & $\Delta\lambda$ \\
\hline
O VI & 1031.91        & 6.9 &0.11             	& 3.6 &0.13						  &423.4  &0.11 \\
O VI & 1037.61        & 3.3 &0.12             	& 1.8 &0.13 						&163.8  &0.10 \\
C II & 1037.02        & 1.6 &0.09             	& 0.8 &0.10 						&1.6    &0.08 \\
C II & 1036.34        & 1.3 &0.09             	& 0.7 &0.10 						&1.5    &0.08 \\
\hline
\end{tabular}
\end{center}
\label{tb1}
\end{table}

A more reasonable estimate is $\pm15\%$ based on comparisons of SUMER and Solar-Stellar Irradiance Comparison Experiment (SOLSTICE) observations \citep{wil1999}. Additional uncertainty in the sunspot O VI intensities arise due to the count rate exceeding the instrumental capabilities, as discussed by \citet{cur2001}. This saturation leads to a 17\% underestimation in intensity for the O VI lines. Uncertainties in radiometric calibration do not propagate fully to the calculated uncertainties of line intensity ratios. A 10\% uncertainty in the O VI 1032/1038 intensity ratio may be a reasonable estimate (N. Labrosse, 2005, \textit{private communication}). This is likely to be a minimum uncertainty for the O VI intensity ratios measured in the sunspot observation due to the saturation.

\begin{figure}
\centering
\includegraphics[width=0.42\textwidth]{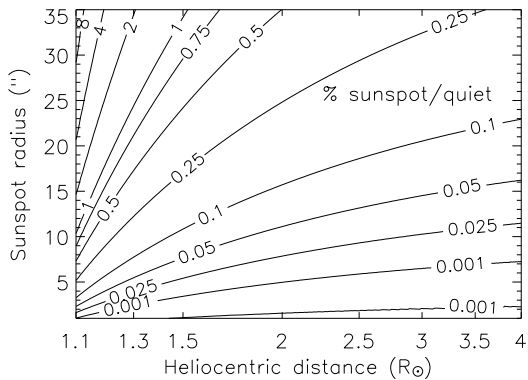}
		\caption{The percentage ratio of the solid angle subtended by a circular sunspot over the solid angle subtended by the rest of the solar disk as a function of height in the corona and sunspot radius, calculated for a sunspot centered on the disk center.}
\label{f2}
\end{figure}

Three points of relevance to coronal resonance scattering emerge from the analysis of disk spectra:
\begin{enumerate}
	\item The O VI 1032/1038 line intensity ratio in quiet Sun and coronal hole regions is close to 2, as expected from theoretical consideration in a collisional plasma. From table \ref{tb1}, the same ratio in sunspots is 2.6 for the peak intensities, and 2.8 for the line integrated intensities. In simple terms, static O$^{5+}$ ions should give a O VI 1032/1038 line intensity ratio close to 4 in the corona, given a collisional exciting disk radiation O VI intensity ratio of 2 \citep{koh1982}. An increase of the O VI disk intensity ratio leads to an increase in the intensity ratio measured in the corona at lower outflow velocities. The enhancement of the 1032 line relative to the 1038 line, giving a line intensity ratio significantly larger than 2, is likely due to radiative excitation of chromospheric O$^{5+}$ ions by a molecular hydrogen (H$_2$) line emitted from the cool sunspot. The 1-1 band of the H$_2$ Q3 transition is blended with the O VI 1032 line \citep{bar1979}. Analysis of the H$_2$ lines observed by SUMER in the 1999 March 18 sunspot is made by \citet{sch1999} but no detailed analysis of the interaction between the emissions of H$_2$ and chromospheric O$^{5+}$ is made.
	\item The O VI lines observed over sunspots are a factor of $\sim$60 more intense than those observed in the quiet Sun. Therefore even small sunspots can give a non-negligible contribution to the total disk radiation seen by coronal O$^{5+}$ ions at lower heights in the corona. A basic factor is the ratio of the solid angle subtended by a sunspot over the solid angle subtended by the rest of the solar disk, as seen from a point in the corona. Figure \ref{f2} plots this ratio as a function of sunspot size and height in the corona for the simple case when the position of the sunspot on the solar disk is directly on disk center as seen from the corona.  Appreciable H$_2$ emission is restricted to sunspots \citep{jor1978} therefore enhanced O VI intensity ratio in the chromosphere, to values significantly above 2, is also restricted to areas that are illuminated by the sunspot, and is likely to be somewhat larger than the sunspot area. The area of the enhanced O VI intensity ratio is an important factor which will be considered in depth in a following study. Under the assumption that the area of enhanced O VI intensity ratio is restricted to the sunspot area, the ratios shown in figure \ref{f2} can be multiplied by the factor of 60 to give an approximate value to the weighting of sunspot to quiet Sun O VI intensity. Even large sunspots have little effect on the behavior of the coronal O VI intensity ratio at heights above $\sim$3$R_\odot$. Limb brightening is not included in the calculations and would decrease the sunspot weighting. Coronal holes would increase the sunspot weighting due to their lower O VI intensity.
	\item The C II lines from the quiet Sun and sunspot regions are very similar. At higher coronal outflow velocities the O$^{5+}$ ions lose resonance with the disk O VI lines and the 1038 coronal line resonates with the C II disk lines while the 1032 line loses its radiative component. Since the C II lines from sunspots are largely unchanged from the quiet Sun lines, sunspots have little effect on the behavior of the coronal O VI intensity ratio at higher outflow velocities. The enhancement of O VI intensity in sunspots while the C II lines remain the same is in agreement with observations made by the Harvard spectrometer on the Apollo Telescope Mount \citep{fou1974}, and can be understood in terms of the different formation temperatures of the ions.
\end{enumerate}

\section{SIMPLE CORONAL MODELS}
\label{model}

The behavior of the coronal O VI 1032/1038 intensity ratio in the presence of sunspots is investigated here using two simple models of the corona - a coronal hole and a streamer. For each model, a line of sight (LOS) with a plane of sky (POS) height of $1.8R_\odot$ is chosen and the O VI line profiles are calculated using the standard radiative and collisional equations \citep{wit1982, li1998}. In the model, the Sun is not approximated as a point source of radiation. The electron temperature $T_e$, of importance in calculating the collisional excitation rates, is 1 MK at all heights for both models. The ion velocity distribution is Maxwellian in the calculations using isotropic temperature distributions, and biMaxwellian otherwise. Ion perpendicular temperature, $T_\bot$, is the temperature perpendicular to the solar radial direction. $T_\bot$ is 6 MK in the coronal hole and 2.7 MK in the streamer. These are values consistent with O$^{5+}$ effective temperatures calculated from O VI linewidths measured by UVCS \citep{mor2005}. $T_\bot$ is kept constant along the LOS. The calculations have been made for two ion temperature anisotropies - an isotropic distribution ($T_\| = T_\bot$) and an anisotropic distribution ($T_\| = T_e$). The O VI intensity ratio is calculated for outflow velocities between 0 and 300 km s$^{-1}$, which are kept constant along the LOS for each calculation.

The spherically symmetric coronal hole model has a radial electron density profile given by

\begin{equation}
n_e = \frac{1\times10^8}{r^8} + \frac{2.5\times10^3}{r^4} + \frac{2.9\times10^5}{r^2}
\end{equation}

\noindent
where $r$ is the heliocentric height in solar radii and the resulting density is in electrons per cm$^{3}$ \citep{doy1999}. The streamer electron density is given by 

\begin{equation}
n_e = \frac{3.60\times10^8}{r^{15.3}} + \frac{9.90\times10^7}{r^{7.34}} + \frac{3.65\times10^7}{r^{4.31}}
\end{equation}

\noindent
\citep{gib1999}. Therefore the POS electron density is $10^{6}$ cm$^{-3}$ for the coronal hole and $4.3\times10^{6}$ cm$^{-3}$ for the streamer at $1.8R_\odot$. Ion density is calculated from the electron density using the ionization equilibria of \citet{maz1998} and the oxygen abundances of \citet{fel1998}, both tabulated in the Chianti atomic database. The coronal hole calculations are made along a LOS that extends to heights where the O VI emission becomes negligible compared to the emission at the POS. The streamer calculations are made along a curtailed LOS which extends $0.25R_\odot$ in each direction along the POS. These are simplistic representations of coronal hole and streamer geometries. 

The synthesized O VI lines are fitted to Gaussians and the 1032/1038 intensity ratio is calculated from the integrated line intensities. The calculations are made for both a quiet Sun disk spectrum and a disk with a sunspot contribution. This contribution is modeled as a large sunspot of radius 20" centered at the point on the solar disk directly beneath the coronal LOS.

\begin{figure}
\centering
\includegraphics[width=0.5\textwidth]{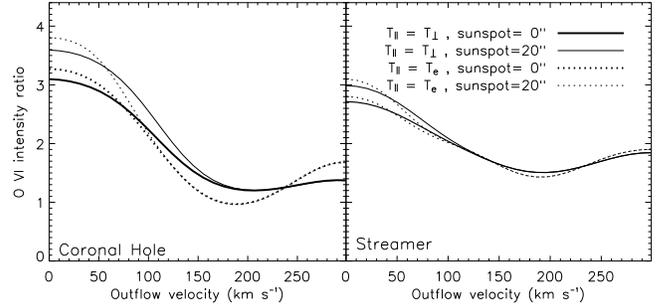}
		\caption{The behavior of the coronal O VI 1032/1038 intensity ratio as a function of outflow velocity with and without a sunspot contribution to the disk radiation. This is calculated here at a height of $1.8R_\odot$ for two coronal models - a coronal hole (left) and a streamer (right). For both models, the intensity ratio has been calculated for two temperature anisotropies as shown in the legend.}
\label{f3}
\end{figure}

Results are shown in figure \ref{f3}. The streamer intensity ratio profiles as a function of outflow velocity are less exaggerated in comparison to the coronal hole profiles, that is the streamer profiles vary less from the collisional intensity ratio value of 2 due to the higher streamer electron density. As expected, the intensity ratios calculated using isotropic temperature distributions show less sensitivity to the effects of Doppler dimming and pumping with outflow velocity. For low outflow velocity, all the profiles show a considerable increase in intensity ratio with the inclusion of a sunspot contribution in the disk spectrum. With an anisotropic temperature distribution, this increase is only seen for velocities below $\sim$100 km s$^{-1}$. The increase continues to higher outflow velocities with an isotropic temperature distribution. This is due to the low $T_\|$ of the anisotropic distributions losing resonance with the disk O VI lines at lower outflow velocities than the high $T_\|$ of the isotropic distributions.

\section{CONCLUSIONS AND PERSPECTIVES}
\label{conclusions}

The analysis and semi-empirical modeling of observed O VI coronal lines leads to constraints on O$^{5+}$ ion parameters. In general, there are large uncertainties in the final estimates of outflow velocity. This is mainly due to the lack of an independent observational constraint on $T_\|$, so that the inferred outflow velocity is a function of $T_\|$. Large uncertainties also arise from the distribution of structures along the line of sight, the flow direction of the solar wind, and the basic uncertainty in measurement, particularly in low intensity coronal holes. At lower coronal outflow velocities, this letter shows that the presence of sunspots on the Sun's disk can lead to a considerable increase in the coronal O VI intensity ratio and should not be neglected in the interpretation of coronal O VI observations. Until more is understood about the nature of O VI and H$_2$ emission above sunspots, this is an additional uncertainty in the determination of O$^{5+}$ ion parameters. Empirical models of O VI observations which neglect sunspots on the disk give an underestimation of the outflow velocity of coronal O$^{5+}$ ions. This underestimation will be worse at lower outflow velocities and will become negligible only at outflow velocities above $\sim$100 km s$^{-1}$, depending on the ion $T_\|$. 

Several observations of a solar minimum west equatorial streamer were made by UVCS during 1997 April 22-27 \citep{hab1997}. \citet{str2002} modeled these observations and found insignificant O$^{5+}$ outflow velocity at heights below $\sim3.5R_\odot$ along the streamer stalk, with an abrupt rise to non-zero outflow velocity above these heights. This was interpreted as evidence of the trapping of coronal plasma by closed magnetic fields in the streamer core. From Michelson Doppler Imager (MDI)/SOHO images, it appears that sunspots were present on the west limb during 1997 April 22. The inclusion of sunspots in the streamer line emission model may lead to non-zero outflow velocity in the streamer core. A re-examination of these observations will be made in future work.

The study presented in this letter is based exclusively on the observations of one sunspot by SUMER on March 18, 1999, associated with active region NOAA 8487 \citep{cur2001}. It would be very desirable to conduct a more detailed study of O VI and C II emission from this and other sunspots, and from active regions in general. This would be essential in interpreting coronal O VI observations correctly. The interaction between molecular hydrogen and chromospheric O$^{5+}$ needs to be understood and is an interesting subject in itself. A simple model is currently being developed and explored with the aim of interpreting the observed O VI 1032 and H$_2$ blended lines above sunspots.

\acknowledgments{{\it Acknowledgments:} Our gratitude to Dr. Xing Li at the University of Wales, Aberystwyth for assistance on coronal O VI modeling, Dr. Nicholas Labrosse at the University of Wales, Aberystwyth for assistance on the SUMER instrument and Dr. Jeff Kuhn at the IfA, Hawaii for his insight on molecular hydrogen. SOHO is a mission of international cooperation between ESA and NASA.}



\end{document}